\documentclass[preprint,nofootinbib,showpacs,superscriptaddress,preprintnumbers,amsmath,amssymb,floatfix]{revtex4-1}

\usepackage{graphicx}
\usepackage{dcolumn}
\usepackage{bm}
\usepackage{color}
\usepackage[usenames,dvipsnames]{xcolor}

\def\Sigmaint{$\Sigma$\kern-.65em\raise.30ex\hbox{$\int$}}
\def\loptwo#1#2{{\ell_{#1}(\ell_{#1}+1)\over#2^2}}

\def\secderv#1{{{\rm d}^2\ \over {\rm d}{#1}^2}}
\def\sumint#1{\sum_{\mathstrut\kern.3em#1}\kern-1.30em\int\kern.60em}

\begin{document}

\title{Quantum-mechanical calculations of cross sections for electron collisions with atoms and molecules}
\author{Klaus~Bartschat}
\email[Electronic address: ]{klaus.bartschat@drake.edu}
\affiliation{Department of Physics and Astronomy, Drake University, Des Moines, Iowa, 50311, USA}
\author{Jonathan~Tennyson}
\email[Electronic address: ]{j.tennyson@ucl.ac.uk}
\affiliation{Department of Physics and Astronomy, University College London, London WC1E 6BT, UK}
\author{Oleg Zatsarinny}
\email[Electronic address: ]{oleg.zatsarinny@drake.edu}
\affiliation{Department of Physics and Astronomy, Drake University, Des Moines, Iowa, 50311, USA}

\date{\today}

\begin{abstract}

  An overview of quantum-mechanical methods to generate cross-section
  data for electron collisions with atoms and molecules is presented.
  Particular emphasis is placed on the time-independent close-coupling
  approach, since it is particularly suitable for low-energy
  collisions and also allows for systematic improvements as well as
  uncertainty estimates.  The basic ideas are illustrated with
  examples for electron collisions with argon atoms and methane.
  For many atomic systems, such as e-Ar collisions, highly reliable
  cross sections can now be computed with quantified uncertainties.
  On the other hand, while electron collision calculations with molecules 
  do provide key input data for plasma models, the 
  methods and computer codes presently used require further development to make 
  these inputs robust.

\end{abstract}


\maketitle

\section{Introduction}\label{sec:Intro}
\label{sec:Introduction}
Electron collisions with atoms, ions, and molecules are well known to
be of critical importance in the understanding and the modeling of
laboratory and technological plasmas, astrophysical processes, lasers,
and planetary atmospheres --- to name just a few examples. Since the
principal motivation of benchmark experiments in this field has been
the test of theoretical models in as much detail as possible, without
averaging, for example, over unpolarized initial beams and summing or
integrating over unresolved observables in the final state, many of
these experiments produce data that are differential in energy, angle,
and even the spin of the particles involved.  On the other hand,
plasma modelers often need angle-integrated cross sections 
over extended energy ranges to
calculate the relevant rate coefficients, sometimes resolved for
particular initial and final states and sometimes just lumped over a
number of possibilities.  Of particular importance, however, is the
need for {\it absolute\/} cross sections, which are notoriously
difficult to obtain in experimental setups, especially in
state-to-state resolved transitions involving neutral targets.  Some
information and additional references can be found in~\cite{Wiley}.
   
Furthermore, state-of-the art plasma modeling often requires data for
a very large number of atomic and molecular processes~\cite{PNAS2016}.
It is virtually impossible to determine all the quantities needed by
experimental means. The reason is not just the cost involved, but
rather the fact that certain cross sections, which may be important in
simulations of particular plasma environments, are not accessible to
standard crossed-beam or gas-cell setups.  Examples include
transitions between excited states, especially if the initial state is
not even metastable, and essentially all collisional data
involving open-shell molecules, which are generally known as radicals.  
As a result, plasma modelers increasingly rely
on theoretical predictions, rather than on educated guesses for the
relevant cross sections and rate coefficients.

In light of the enormous progress made in both computational hardware
and software, it has indeed become possible to generate a large amount
of the required data by numerical means, and the results are typically
collected in databases.  Some information can be found
in~\cite{0953-4075-49-7-074003,LXCat-PPaP}.  Interestingly, the concept of uncertainty estimates,
i.e., an assessment of the reliability of particular theoretical
predictions, has only recently drawn significant attention among both
data users and data producers.  While the publishers of experimental
data are routinely expected to provide some uncertainly estimates (as
difficult as this may be for systematic errors), this has not been a
widely practiced tradition for theorists~\cite{PhysRevEds}.

The present manuscript is organized as follows.  We begin with a brief
overview of quantum-mechanical methods developed for electron
collisions with atoms and molecules, before concentrating on the
time-independent close-coupling (CC) approach, which has been
extensively used to obtain data for low-energy and more recently also
for intermediate-energy collisions.  After illustrating the basic
ideas for electron-atom scattering, the extensions needed to handle
molecular targets are outlined. The effect of various approximations,
their advantages and shortcomings, are then illustrated for electron
collisions with argon atoms and methane.  Unless mentioned otherwise,
atomic units are used throughout this paper.

\section{Overview of Numerical Approaches}\label{sec:Numerical}

There is a wealth of literature available on methods for electron
scattering calculations.  Instead of repeating this information in
great detail, only the minimum framework needed for understanding the
basic ideas will be presented here. For a more thorough treatment and
overviews, standard textbooks on collision theory and recent
reviews~\cite{Madison-AlHagan,jt474,Burke2011,Bray2012135,0022-3727-46-33-334004}
should be consulted. Some descriptions and example computer programs for atomic structure and collisions 
can be found in~\cite{Bartschat1996}.
In this paper, we will illustrate the basic
principles on a few concrete examples that are well known to be
important for plasma applications.

Unfortunately, there is no unambiguous rule regarding the
reliability of a particular method.  Nevertheless, as will be discussed 
further below, the collision energy is
an important parameter that can provide some general guidelines.
More details, in particular with regard to estimating the uncertainty and the reliability of
theoretical predictions, can be found in~\cite{jt642}.

Numerical approaches for atomic collision processes, and in particular for electron-atom
and electron-molecule collisions, are often classified as perturbative or non-perturbative,
as well as time-dependent or time-independent.  
Time-dependent formulations are based on the time-dependent Schr\"odinger equation (TDSE),
\begin{equation}\label{eq:TDSE}
i \frac{\partial}{\partial t} |\Psi(\bm{X},t) \rangle = H(\bm{X},\bm{P},t) |\Psi(\bm{X},t) \rangle.
\end{equation}
for the wave\-function~$|\Psi(\bm{X},t) \rangle$, or -- in a fully relativistic framework -- on the corresponding Dirac equation. 
Here $\bm{X}$ collective denotes all spatial ($\bm{r}_i,~i=1, ..., N+1$) coordinates and spin projections ($\sigma_i$) of
the $N$ target electrons plus the projectile. The operator $H$ is the Hamiltonian,
containing the kinetic energy~$K$ of the particles and their potential energy~$V$ in the field of the
target nucleus as well as their mutual interaction. If written in coordinate space,
the momentum operator~$\bm{P}$ (again representing collectively the linear momenta of all $N+1$ electrons) acts as
a derivative operator with respect to the coordinates.
In principle, $H$ may contain an explicit time dependence
(for example, in short-pulse laser-atom/molecule interactions), but 
here we will concentrate on steady-state scenarios and only consider the 
time-independent kinetic and potential energies mentioned above.

The TDSE is a partial-differential equation that can be solved as an initial-value problem by propagating 
a known (prepared) initial state, usually the product of a 
wave\-packet for the projectile and the initial state of the target, in time until the collision process is
finished.  In the ``time-dependent close-coupling'' (TDCC)~\cite{PhysRevA.65.042721} approach, the ($N+1)$-electron
wave\-function is expanded in some suitable form (see also below), and
the relevant information is extracted from the propagated solution using a number
of different techniques. While TDCC has been successful for some benchmark studies, particularly 
in the description of electron-impact ionization processes,
it is not widely used for production calculations of atomic or molecular data.  Hence we will not consider
it further in this paper.

Since the collision processes of interest here are steady-state scenarios, the calculations
can also be performed using a time-independent formulation.  This requires the solution of the
time-independent Schr\"odinger equation (TISE),
\begin{equation}
H(\bm{X},\bm{P}) |\psi_E(\bm{X}) \rangle = E |\psi_E(\bm{X})\rangle 
\end{equation}
for the wave\-function of the collision system for a fixed total
energy~$E$, subject to particular boundary conditions for the
projectile and the target. For ionization processes in particular,
these boundary conditions are very complex.  As a result, the proper
matching in the asymptotic region is a major challenge, when two (or
even more) free electrons are far away from the charged residual
target ion.  While this challenge has been addressed, for example, by
the technique of ``exterior complex scaling''
(ECS)~\cite{Rescigno24121999}, that method has not been used
extensively for production calculations either.

It is also possible to formulate the problem in momentum space.  In
this case one solves the Lippmann-Schwinger (LS) equation for the
transition matrix, from which all observable parameters, such as cross
sections, for the collision can be derived.  One of several ways to
express the LS equation is~\cite{McCarthyWeigold}
\begin{equation}\label{eq:LS1}
\langle \bm{k}_f \Phi_f | T | \bm{k}_i \Phi_i \rangle = \langle \bm{k}_f \Phi_f | V | \Psi_i^+ \rangle .
\end{equation}
Here $|\bm{k}_i \Phi_i \rangle$ and $|\bm{k}_f \Phi_f \rangle$  are the initial and final asymptotic states, namely
plane or Coulomb waves for the projectile with initial (final) linear momentum \hbox{$\bm{k}_i~(\bm{k}_f)$} and 
the corresponding $N$-electron target states \hbox{$\Phi_f~(\Phi_f)$}, while
$T$ is the transition operator.  Alternatively, $T$ can be replaced by the
simpler interaction operator~$V$ if the exact solution of the scattering problem (here denoted as $|\Psi_i^+ \rangle$)
is used instead.  The formal solution for~$|\Psi_i^+ \rangle$ is given by
\begin{equation}\label{eq:LS2}
|\Psi_i^+ \rangle  = | \bm{k}_i \Phi_i \rangle + { 1 \over E^{(+)} - K} \,V\, | \Psi_i^+ \rangle
\end{equation}
where $K$ is the kinetic-energy operator and the superscript ``$(+)$'' indicates the appropriate boundary conditions.

Both the TDSE in coordinate space and the LS equation can only be
solved exactly for model problems.  The LS formulation, however, not
only allows for a non-perturbative treatment using the ideas of
``convergent close-coupling'' (CCC)~\cite{Bray2012135} in momentum
space, but also for a systematic derivation and classification of the
above-mentioned perturbative methods.  Equation~(\ref{eq:LS2}) in
particular, provides an iteration scheme for the determination
of~$|\Psi_i^+ \rangle$.  Starting with just the first term on the
right-hand side, one obtains the first-order ``Plane-Wave (or
Coulomb-Wave) Born Approximation'' (PWBA or CWBA), in which the
projectile-target interaction is treated as a perturbation, i.e., one
only needs the matrix elements of this interaction between otherwise
unperturbed zero-order wave\-functions.  Higher-order versions of this
formulation can also be derived, and the method can be further
improved by treating part of the interaction more accurately through
the use of ``distorted'' rather than plane waves, thereby resulting in
variants of the ``Distorted-Wave Born Approximation'' (DWBA).
Unfortunately, it is difficult to estimate the uncertainty of
theoretical predictions based on these perturbative formulations,
since higher-order terms are difficult to calculate.  Hence their
importance often remains unknown --- unless comparison with experiment
or results from reliable non-perturbative calculations are available.

For completeness, we also mention the existence of semi-empirical
methods, such as the ``Binary Encounter f-scaling''
(BEf)~\cite{PhysRevA.64.032713} and ``Binary Encounter Bethe''
(BEB)~\cite{PhysRevA.50.3954} approaches to electron-impact excitation
and ionization.  While these methods are useful in practice, they are
somewhat limited in scope.  For example, BEf can only be used for
optically allowed transitions and also requires experimental or
reliable theoretical data for rescaling.  Once again, it appears
practically impossible to suggest a method for assigning an
uncertainty associated with these approaches other than by benchmarking
against experiment.

In the next section, we will concentrate on just one method, namely
time-independent close-coupling formulated in coordinate space. This
is a non-perturbative approach that, in principle, allows for the
solution of the TISE.  In practice, of course, the method has limitations
as well, both regarding the basic formulation (e.g., how to include
the proper boundary condition for ionization processes), the physics
accounted for (e.g., the neglect or approximate treatment of
relativistic effects), and numerical aspects (e.g., the discretization
of derivatives and integrals).  These aspects will be discussed in
some detail below.
 
\section{Time-independent close-coupling in coordinate space}

Time-independent close-coupling has been the method of choice for
treating low-energy electron collisions for many years.  It is based
upon an expansion of the total wavefunction for a collision system in
terms of a sum of products that are constructed from $N$-electron
target states~$\Phi_i$ and functions~$F_{E,i}$ describing the motion
of the projectile for a total (target $+$ projectile) collision
energy~$E$.  If rela\-tivistic effects are neglected, the total
symmetry of the scattering system, $\Gamma$, comprises a total
spin~$S$ and a parity~$\pi$; for atoms the orbital angular
momentum~$L$ is also conserved while for molecules this property
depends on the (point-group) symmetry of the molecule.  The
wavefunction for each $\Gamma$ is, in its general form, expanded as
\begin{eqnarray}\label{eq:CCexpansion}
 \Psi^\Gamma(\bm{x}_1,\ldots,\bm{x}_{N},\bm{x}) & = & \,
 {\cal A\/}  \, \sumint i \sum_j a_{i,j}^\Gamma
\Phi_i^\Gamma(\bm{x}_1,\ldots,\bm{x}_N,\hat{\bm{x}}) 
       \, {1 \over r} \, F_{E,i,j}(r) \nonumber \\
       & & ~~~ +   \sum_k b_k^\Gamma \chi_k^\Gamma(\bm{x}_1,\ldots,\bm{x}_N, \bm{x}_{N+1})\,.
\end{eqnarray}
This results in symmetry-dependent ``partial waves'' that have to be summed
until convergence is reached.  In Eq.~(\ref{eq:CCexpansion}), 
\Sigmaint\ denotes a sum over all discrete and an integral over all continuum
states of the target, and ${\cal A}$ is the anti\-symmetrization
operator that accounts for the indistinguishability of the projectile
and the target electrons.  Furthermore, the angular and spin
coordinates of the projectile electron have been coupled with the
target states to produce the ``channel functions''
$\Phi_i^{\Gamma}(\bm{x}_1,\ldots,\bm{x}_N,\hat{\bm{x}})$.  The second
sum in the first term is needed because  some atomic and all
molecular collisions require the consideration of more than one set of
asymptotic functions for a given target state and overall symmetry.
The second term in Eq.~(\ref{eq:CCexpansion}) contains wavefunctions where
the scattering electron is placed in orbitals associated with the
target; this term is sometimes called ``$L^2$'' as it only contains
localized functions which are $L^2$-integrable. The $L^2$ term is not
required if the first \Sigmaint\ term is complete, but it is actually a
feature of many practical implementations of CC procedures.  The
unknown coefficients $a_{i,j}^\Gamma$ and $b_k^\Gamma$ are determined
in the calculation; in the \hbox{$R$-matrix} method~\cite{Burke2011,jt474}, for example, 
these are the eigenvector
coefficients obtained from diagonalizing an (inner region) Hamiltonian
matrix.
\par
The target states~$\Phi_i$ are usually chosen as multi-configuration expansions.  They
are generally not eigenstates of the $N$-electron target Hamiltonian, but rather 
diagonalize it according to
\begin{equation}\label{eq:target}
\langle \Phi_{i'}|H_T^N|\Phi_{i}\rangle = E_i \delta_{i'i}.
\end{equation}
As will be further outlined below, this diagonalization property can be very useful in 
generating so-called ``pseudo-states''.

For atomic systems, the radial wavefunctions $F_{E,i}(r)$ for
the projectile are determined from the solution of a system of coupled
integro-differential equations given by
\begin{equation} \label{eq:CCpots}
\left[ \secderv r - \loptwo ir + k^2 \right] \, F_{E,i}(r)  = 2 \sumint j
V_{ij}(r) \, F_{E,j}(r) + 2 \sumint j W_{ij} \, F_{E,j}(r),
\end{equation}
with the direct coupling potentials 
$V_{ij}(r)$ and the exchange terms
$W_{ij}F_{E,j}(r)$.  While it is difficult to write out these potentials explicitly,
they can be evaluated by existing computer programs. For molecules
equations of this form are used to generate a suitable set of basis
functions with which to represent these functions~\cite{jt286}.  
In this case, however, only very simplified potentials are usually employed.
\par
The collision problem essentially consists of finding the solution to this system
of coupled equations for each total energy subject to the appropriate boundary conditions.
This can be achieved by various
iterative, non-iterative, or algebraic methods.  A frequently used approach is
the $R$-matrix method developed by Burke and collaborators in Belfast over the past
few decades. A comprehensive summary of a large variety of $R$-matrix applications
has been given by Burke~\cite{Burke2011}.  Also, a suite of computer codes for atomic and ionic targets 
was published~\cite{BEN95} and updates are available, for example, from Badnell's website~\cite{Badnell-web}. 
Similarly, the UK Molecular \hbox{$R$-matrix}
codes have been routinely updated and published over time \hbox{\cite{jt161,jt225,jt518}}. The current implementation, 
known as UKRMol~\cite{jt518}, is obtainable via the open-access CCPForge program repository and can be run through
the Quantemol-N expert system~\cite{jt416}. An updated version, known
as UKRMol+, which copies the latest atomic codes and
uses \hbox{$B$-spline} basis functions to represent the radial
part of the channel functions, is currently being developed~\cite{jtaddref}. 

When inserted into the 
non-relativistic TISE, Eq.~(\ref{eq:CCexpansion}) should allow for an accurate solution.  In practice, however,
the sum over the (countable) infinite number of target bound states and the
integral over the (uncountable) ionization continuum cannot be carried out
by any computational implementation. In the following, we will assume that 
purely numerical aspects, such as the discretization of the radial grid in the
evaluation of derivatives and integrals, can be sufficiently well controlled that
errors are practically negligible compared to the approximations made regarding the underlying physics.  
The principal issue of concern, therefore, is the
approximate treatment of these infinite sums.  

It turns out that the collision energy is an important, though
unfortunately not the only, parameter that can be used to construct
sensible approximations from the close-coupling expansion.  For
elastic collisions, a one-state CC expansion provides a start. In
other words, only the state of interest (generally the ground state,
although this is not formally required) is kept in the CC expansion.
When written out explicitly, this \hbox{CC-1} model (where the ``1''
indicates the number of states kept in the CC expansion) leads to an
integro-differential equation for the projectile wave\-function, due
to the non-local character of the exchange term~$W_{11}$ in
Eq.~(\ref{eq:CCpots}).  This is called the ``static exchange'' (SE)
approximation.  Sometimes the non-local term is neglected or replaced
by local approximations, thereby reducing the problem to so-called
``potential scattering''.

As will be discussed below, even when elastic scattering is the only
``open'' channel, i.e., the projectile does not have enough energy to
produce any excitation in the target, CC-1 is often not sufficiently
accurate. This is due to the fact that the charged projectile can
polarize the charge cloud of the target.  If the target does not have
a permanent dipole moment, this leads to an asymptotically attractive potential,
which is proportional to the electric dipole polarizability and falls
off as $1/r^4$ at large distances from the target center. While higher
multipole terms may also enter, this is the leading term.  The CC
expansion can account for this effect by including {\it all\/} states
(energetically open or closed) that can be reached by optically
allowed transitions from any given state.  This immediately brings the
problem back to an infinite expansion, and hence further
simplifications must be sought.  These are:

\begin{itemize}\setlength\itemsep{0em}
\item For some targets, especially the alkali atoms Li, Na, K, \ldots, the 
      coupling of the ground state to the first optically excited state (e.g., $3s - 3p$ in Na) is
      so strong that a \hbox{CC-2} expansion with just these two states can give good results for 
      elastic scattering.  
      Furthermore, a \hbox{CC-4} expansion, also including the $4s$ and $3d$ states,
      is generally appropriate for excitation of the $3p$ state~\cite{0022-3700-5-8-015}.
\item For molecules, an extension of the SE approximation, which is still widely used  
      in studies of  elastic scattering,
      is the ``static exchange plus polarization'' (SEP)
approximation.  In terms of Eq.~(\ref{eq:CCpots}), this approximation
uses a \hbox{CC-1} model augmented by the  $L^2$ terms obtained
by performing single excitations from the target wavefunction.
The SEP model has proved to be rather effective for obtaining converged
results for low-lying shape resonances~\cite{vrs08,jt533}.
\item If the convergence of the CC expansion is very slow, and a large portion of the dipole polarizability
      even comes from coupling to the ionization continuum~\cite{jt468}, a pseudo-state
      may be constructed that reproduces the desired polarizability.  Such \hbox{CC-2-pol}
      models can be very successful in low-energy elastic electron collisions, for example, with noble gases.
      An example will be shown in the next section.
\item Often the non-local character of the polarization potential is
      simplified again by defining a local approximation for this potential, thereby reducing the
      problem once again to simple potential scattering.  In some cases, 
      imaginary ``optical potentials''~\cite{PhysRevA.28.2740} are used to account for the
      possibility of excitation and ionization processes, i.e., loss of flux from the elastic channel.
\end{itemize}

Another important energy range is the near-threshold regime, which is
often affected, or even dominated, by resonances.  In this case the CC expansion
needs to contain a number of $n$~discrete states, and hence the resulting models will be referred to as \hbox{``\hbox{CC-$n$}''}.  
This approach has been the method of choice for many years.  It is indeed often highly
successful in the description of resonances associated with low-lying
inelastic thresholds.  Following up on the discussion above, however, 
the method may have problems in the low-energy
regime if significant polarization effects originate from coupling to
higher-lying discrete states and, in particular, the ionization continuum.
Hence, the \hbox{CC-$n$} method usually does not give results of equal (or even similar) 
quality for all transitions between the states included in the expansion. 

For intermediate-energy collisions, the above-mentioned effect of
coupling to discrete states omitted in the \hbox{CC-$n$} expansion,
and even more importantly to the ionization continuum, should
therefore be accounted for in some way. This can be done by extending
the CC expansion with a number of pseudo-states, which are essentially
finite-range states that are forced to fit into a box. This can be
done by limiting the above-mentioned diagonalization of the target
Hamiltonian to a finite spatial regime, outside of which all target
orbitals must either vanish completely or fall of sufficiently fast
with the distance from the center of the box.  For the basic idea, the
details of the box (``hard'' or ``soft'') are not important; it only
matters that the states are $L^2$-integrable and provide a way to
discretize the (countable) infinite Rydberg and the continuous
ionization spectra.  This is the general principle behind the
``convergent close-coupling'' (CCC)~\cite{Bray2012135} and
``\hbox{$R$-matrix} with Pseudo-States''
(RMPS)~\cite{0953-4075-29-1-015,jt341} approaches.  While the
implementations may vary greatly, the essential idea is exactly the
same in both methods.  Hence, if the same states (physical and pseudo)
are included in the expansion, the final results should be the same.

Recent CCC calculations~\cite{zsf16} for electron-H$_2$ collisions have demonstrated
that this  method indeed provides a complete treatment of the electronic
degrees of freedom over a wide range of energies. However, so far
the CCC methodology has been restricted to the treatment of collisions
with relatively simple, few-electron targets.

For H$_2$ it is possible to get very reliable representations of electronically excited states and
hence electronic impact excitation cross sections~\cite{jt208}, but this is hard for the general
case. Sophisticated quantum chemical methods are available for treating electronically excited states
for small molecules, largely based on procedures such as multi-reference configuration 
interaction~\cite{92KnWexx.ai}. At present, however, these wavefunctions cannot be used directly as the input
for electron-molecule scattering calculations, for several reasons: (a)~the formulation given in Eq.~(\ref{eq:CCexpansion})
requires a single set of orbitals for all states while the best results are obtained with
orbital sets optimised for each state~\cite{jt632}; (b)~it is difficult to create a balanced model
for electron scattering based on an MRCI target wavefunction; (c)~excited states of molecules 
rapidly become very diffuse and are therefore hard to treat, at least within the confines of an
$R$-matrix procedure; and (d)~treatments involving electron collisions with complicated target wavefunctions
rapidly become prohibitively computationally expensive. See Halmov{\'a} {\it et al.}~\cite{jt444} for
an illustrative discussion of this last point.

Moving on to the high-energy regime, we note that the CC expansion can
be linked to perturbative formulations, in that ultimately only the
initial and final (if different) states need to be kept in the CC
expansion. The transition-matrix elements can then be obtained as
integrals between simplified target wavefunctions, using some variants
of the PWBA, CWBA, of DWBA formulations~\cite{Seaton1961}. Such
approximations are also frequently used for partial waves with high
angular momenta, which may be needed to ``top up'' the results to
guarantee convergence of the partial-wave expansion.  The physical
justification for this procedure lies in the fact that the centrifugal
barrier keeps these parts of the projectile wave\-function rather far
away from the target and hence reduces the effects of the complex
short-range interactions.

Finally, other important issues concern the structure description
itself and the way to account for relativistic effects if desired.  As
a rule of some thumb, quasi-one and quasi-two electron systems, such
as the alkali and the alkali-earth elements, are well described by one
or two valence electrons outside of a closed core --- provided, of
course, that one is only interested in transitions involving these
valence electrons.  More complex targets, such as the noble gases
beyond helium and molecules with more than two active electrons, are
much harder to describe, since any excited or ionized state leads to
multiple open shells with non-zero orbital and spin angular momenta.
In general, targets with partially filled sub\-shells can be very
challenging.  Consequently, great care has to be taken to ensure that
the structure description is sufficiently accurate for the subsequent
collision calculation to make sense at all~\cite{jt642}.

Relativistic effects are generally more important
for heavier rather than lighter targets, but the details again depend on whether or not 
the fine-structure is resolved in the process to be described.  Going beyond 
simply neglecting these effects entirely, the following three methods are typically applied:
\begin{itemize}\setlength\itemsep{0em}
\item Recouple results from a non-relativistic model into a relativistic scheme using only
      angular-momentum algebra (Clebsch-Gordan coefficients, Racah symbols, etc.)
\item Include these effects as a first-order perturbation, i.e., calculate matrix elements of 
      operators such as the spin-orbit interaction between properly (re)coupled non-relativistic 
      wave\-functions and include these matrix elements when setting up the interaction matrix.
\item Formulate the entire problem in the Dirac framework.
\end{itemize}
So far only the recoupling approach has been applied to treat spin-orbit
effects in molecules and then only in limited circumstances~\cite{jt531}.
An analogous recoupling  approach is routinely used to treat rotational
excitation in molecules~\cite{jt271}. The evidence is that this
method works rather well~\cite{jt328,jt465-kb}, but in general
rotational distributions are either treated as thermalized or simply
ignored in state-of-the-art plasma models.  Consequently, there has not been
a demand for rotational excitation cross sections from plasma modelers
with the exception of astronomers modelling weakly ionized regions of interstellar medium.
Molecules with permanent dipole moments, however, display very strong
forward scattering, which is hard to measure experimentally. Theory provides
the best means for correcting for this \cite{jt464}.

Without going into further details (some can be found in~\cite{jt642}),
we only mention here that quantum-electric field effects can be safely
neglected in the electron collision calculations of interest for this
paper. Furthermore, if effects due to the hyperfine-structure play a
role (this is most important if the radiation emitted from excited
states is depolarized), this would once again be accounted for by
recoupling only. See~\cite{jt414}, for example.

\section{Illustration: Electron collisions with argon atoms}
We will now illustrate some of the concepts outlined above.  
Since it is a topic of significant interest for plasma physics
and also a good candidate to show the effects of the various approximations, we use exclusively
electron collisions with Ar atoms in this section on electron-atom collisions. A larger variety of
cases can be found in~\cite{0022-3727-46-33-334004} and~\cite{jt642}.

The numerical results presented here were obtained with the $B$-spline $R$-matrix (BSR)
method (see~\cite{0953-4075-46-11-112001} for an overview)
and the associated computer code~\cite{ISI:000235501000004} for solving the
close-coupling equations.  The BSR method has some practical advantages when it comes to
obtaining an accurate target description, due to its ability to employ non-orthogonal
sets of one-electron orbitals to build up the multi-electron states.  As mentioned above and
will be shown with an example below, this is a very important issue in some cases. 
A relativistic version is also available~\cite{ISI:000257288800094}.
Since its 
original publication,  the computer code
has been optimized and parallelized to allow for the use of a large number of pseudo-states.  This
enables the description of ionization processes and also allows for systematic convergence
studies of the CC expansion.  Following the previously introduced naming convention, our models will thus
be labeled \hbox{BSR-$n$} to indicate the number of coupled states.

\subsection{Elastic Scattering and Momentum Transfer}

\begin{figure}[t]
\includegraphics[width=0.60\textwidth]{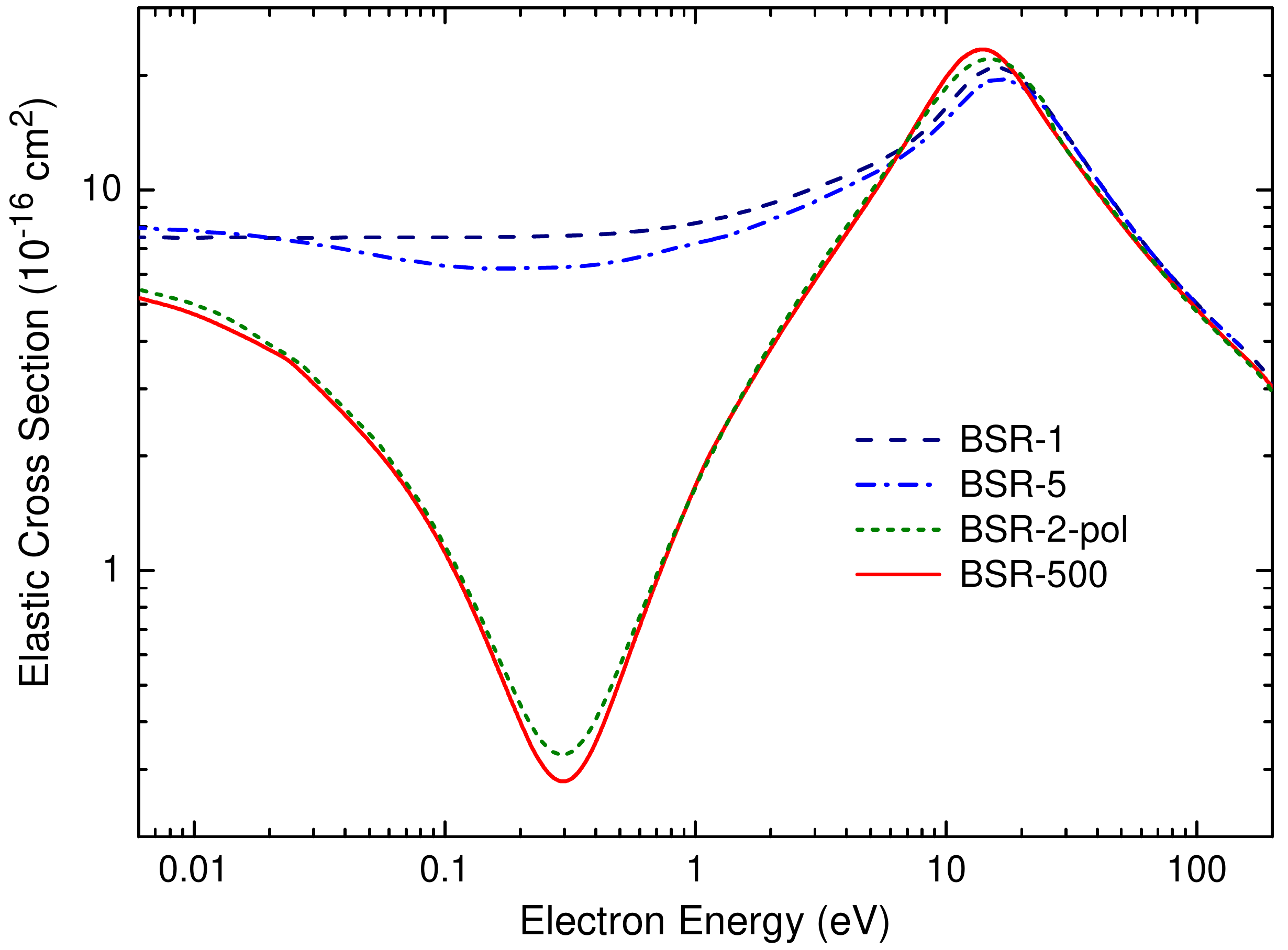}
\caption{BSR results for elastic electron collisions with argon atoms in their ground state,
        as obtained with a varying number of states in the BSR close-coupling expansion. See text for details.}
\label{fig:Ar_els_BSR}
\end{figure}

\begin{figure}[t]
\includegraphics[width=0.60\textwidth]{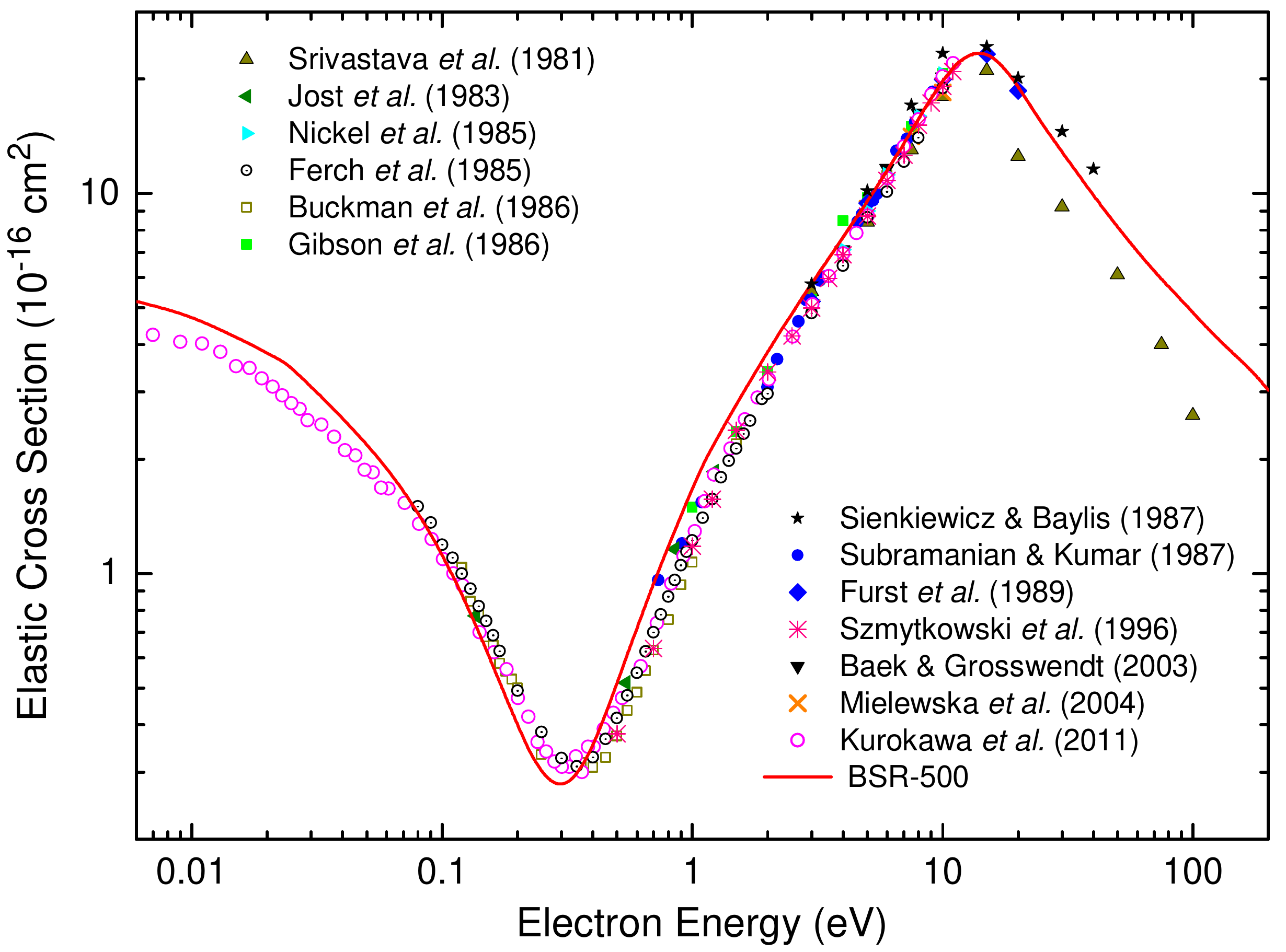}
\caption{Results for elastic electron collisions with argon atoms in their ground state.
         The \hbox{BSR-500} predictions~\cite{BSR500}  are compared with a variety of experimental 
         results~\cite{Srivastava1981,Jost1983,Nickel1985,Ferch1985,Buckman1986,Gibson1986,Sienkiewicz1987,Subramanian1987,Furst1989,Szmytkowski1996,Baek2003,Mielewska2004,Kurokawa2011}.}
\label{fig:Ar_els_comp}
\end{figure}

\begin{figure}[t]
\includegraphics[width=0.60\textwidth]{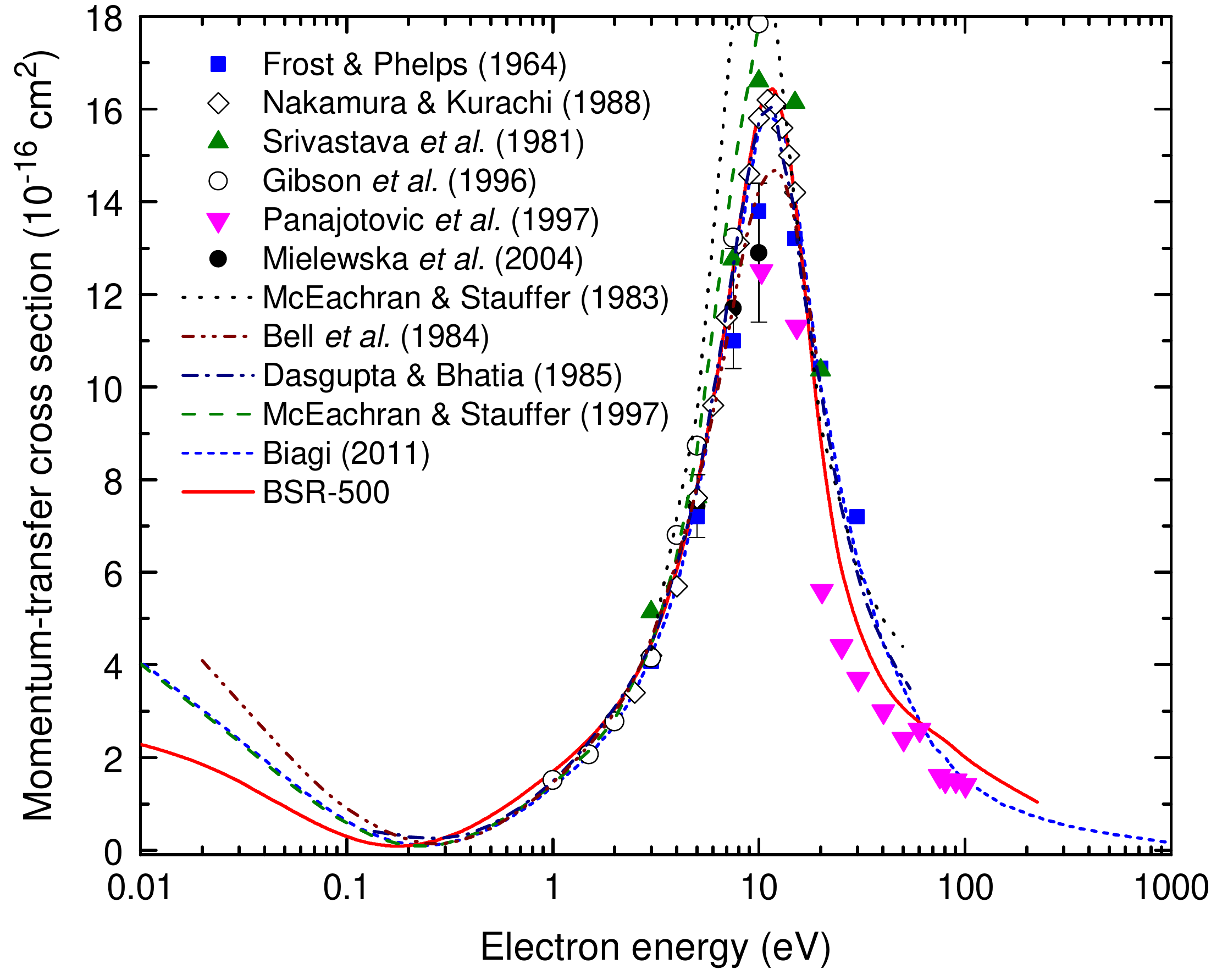}
\caption{Momentum-transfer cross section for argon atoms in their ground state. 
         The \hbox{BSR-500} predictions~\cite{BSR500} are compared with a variety of experimental and other theoretical
         results \hbox{\cite{Frost1964,Nakamura1988,Gibson1986,Panajotovic1997,Mielewska2004,McEachran1983,Bell1984,Dasgupta1985,McEachran1997,Biagi2011}}.}
\label{fig:Ar_mom}
\end{figure}

Figure~\ref{fig:Ar_els_BSR} shows the results for elastic e-Ar collisions, as obtained  
with four different approximations, namely BSR-1, BSR-5, BSR-2-pol, and BSR-500~\cite{BSR500}.
Specifically, these models contain just the ground state (BSR-1), the ground state plus the
lowest four excited states with predominant configuration $3p^5 4s$ (BSR-5), the ground state
plus a single pseudo-state to reproduce the dipole polarizability, and a total of 500 states (BSR-500).
The latter set consists of the lowest 31 physical bound states, 47 pseudo-states with energies 
below the ionization threshold to account for coupling to the high-lying Rydberg states, and 
422 further pseudo-states to simulate the effects of the ionization continuum.  

Looking at the various predictions, we see that the Ramsauer-Townsend minimum is completely missed
by the BSR-1 model and hardly seen in the BSR-5 curve. BSR-2-pol, on the other hand, obtains the basic 
energy dependence of the cross section, with the minimum around an energy of 0.3~eV -- just as the
very extensive \hbox{BSR-500} model. Clearly, a properly constructed single pseudo\-state can handle this particular problem very well.

The accuracy of the fully {\it ab initio\/} \hbox{BSR-500} model is illustrated in 
Figs.~\ref{fig:Ar_els_comp} and~\ref{fig:Ar_mom}, where its predictions are compared
with a number of experimental data for elastic scattering and the momentum-transfer cross section, respectively.
The overall agreement is excellent.

\subsection{Excitation}

\begin{figure}[t]
\includegraphics[height=0.50\textwidth]{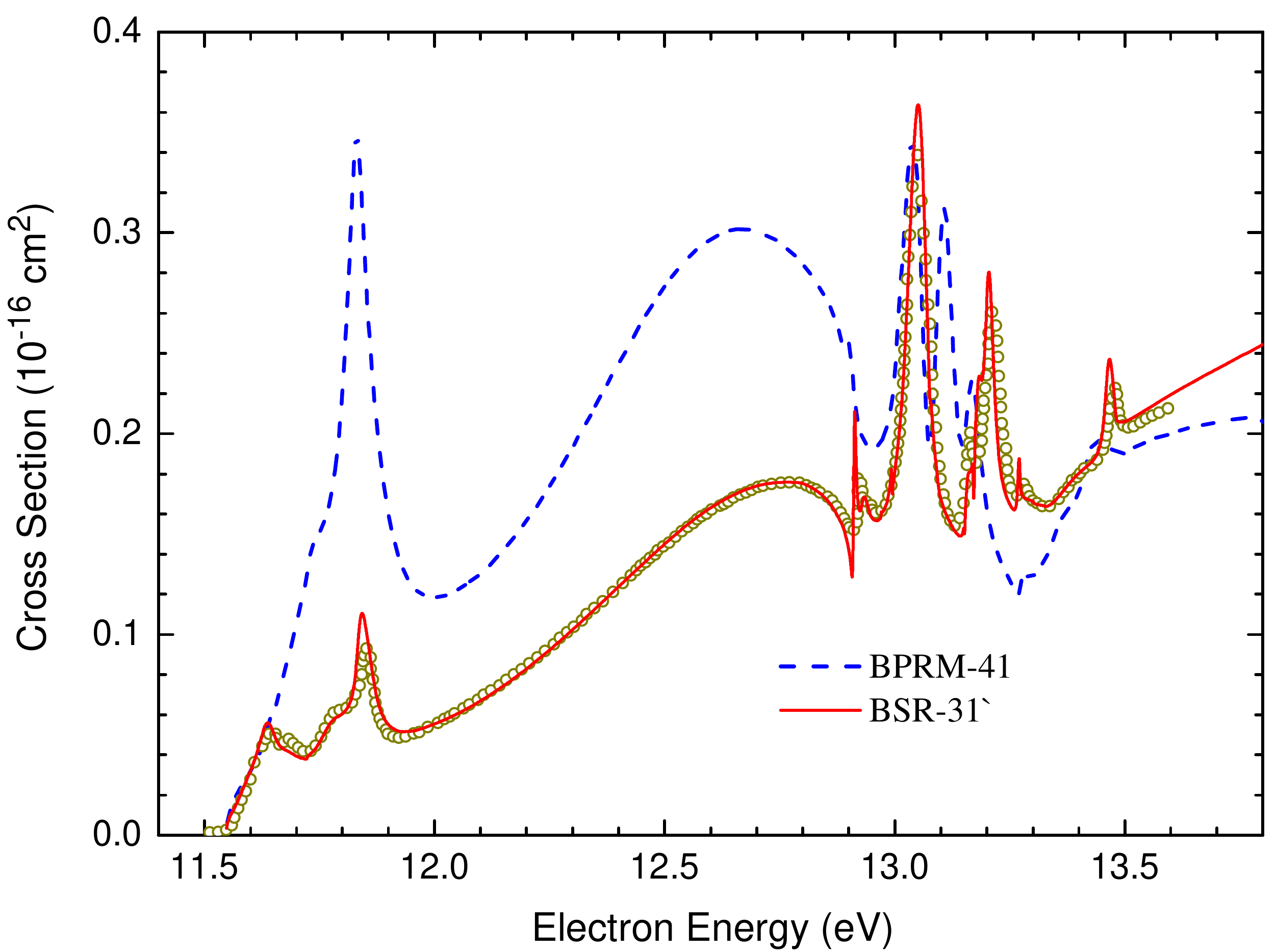}
\caption{Cross sections for electron-impact excitation of the $3p^5 4s$ metastable states 
         (the sum of the $J=2$ and $J=0$ states) in argon from the ground state.
         The experimental data of Buckman {\it et al.}~\cite{0022-3700-16-22-012}, renormalized by a factor of~0.53,
         are compared with predictions from a BSR-31 model~\cite{ISI:000225951700012} and a previous BPRM-41~\cite{PhysRevA.58.1275}
         calculation.}
\label{fig:Ar_meta}
\end{figure}

\begin{figure}[t]
\includegraphics[width=0.70\textwidth]{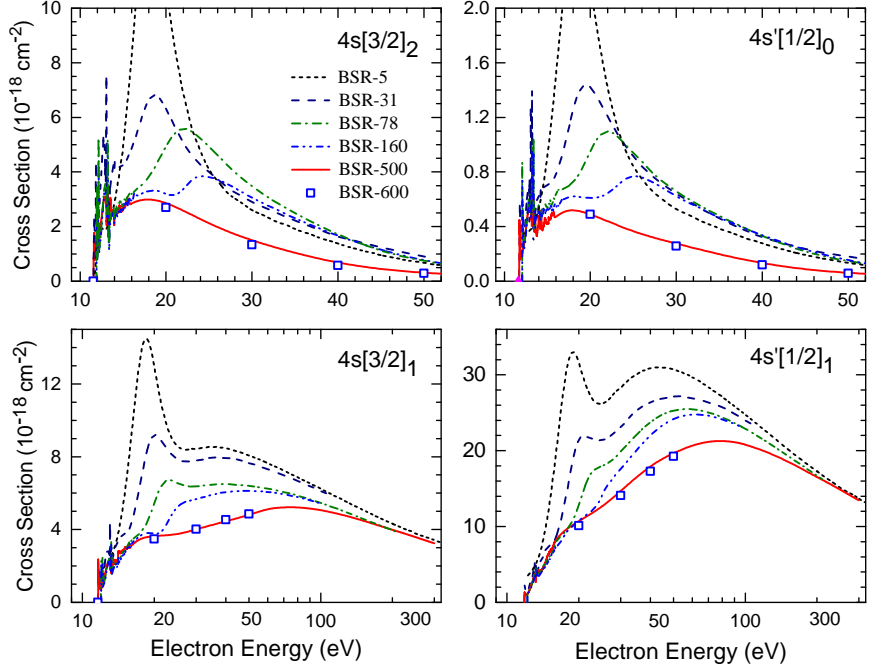}
\caption{\label{fig:Ar-excite} Cross sections for
  electron-impact excitation of the individual states of the $3p^5 4s$ manifold in argon from the
  ground state $(3p^6)^1S_0$.  The results from a number of BSR calculations
  with a varying number of states shows the convergence of the CC expansion.
  }
\label{fig:Ar_convergence}
\end{figure}

\begin{figure}[t]
\includegraphics[width=0.70\textwidth]{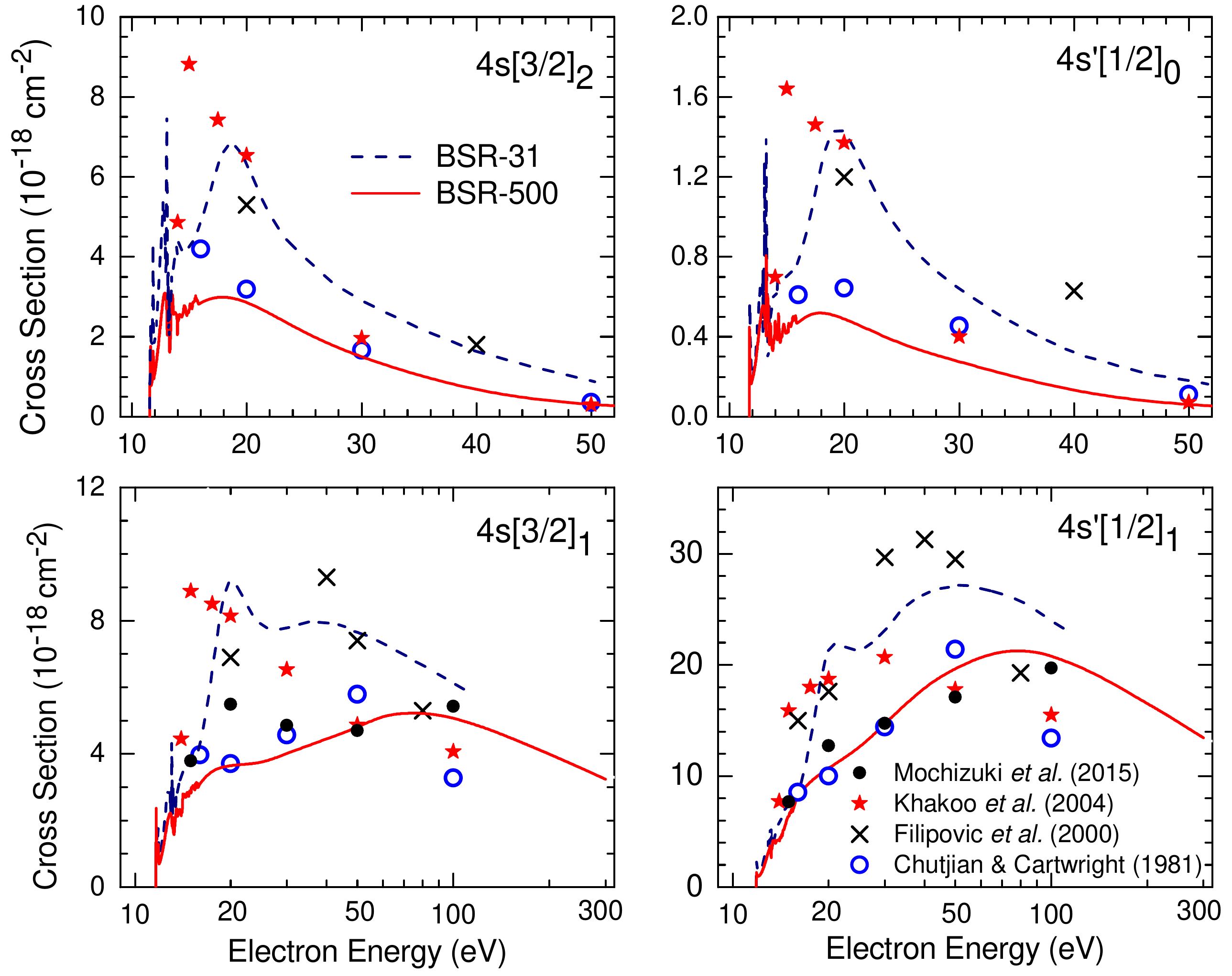}
\caption{Cross sections for electron-impact excitation of the individual states of the $3p^5 4s$ manifold 
         in argon from the ground state.
         The BSR-31 and \hbox{BSR-500} predictions~\cite{BSR500} are compared with a variety of experimental 
         data~\cite{Mochizuki2015,Khakoo2004,Filipovic2000,Chutjian1981}.}
\label{fig:Ar_4s-compare}
\end{figure}

Figure~\ref{fig:Ar_meta} is another example of the impressive progress that the BSR method has achieved in the
description of the near-threshold regime.  The resonance structure in the metastable excitation function is 
reproduced in excellent agreement with the experimental data~\cite{0022-3700-16-22-012}, except for an 
overall renormalization factor of~0.53, which is still within the overall experimental uncertainty.  
The principal reason for the improvement over a previous semi-relativistic Breit-Pauli
\hbox{$R$-matrix} calculation \hbox{(BPRM-41~\cite{PhysRevA.58.1275})} in this case is the
improvement in the description of the target states in this energy region. Although \hbox{BSR-31} is very
 good for
obtaining this
particular cross section, we emphasize that the model would not be not sufficient for either low-energy elastic scattering
(see above) or the intermediate-energy regime discussed next.

\begin{figure}[t]
\includegraphics[width=0.50\textwidth]{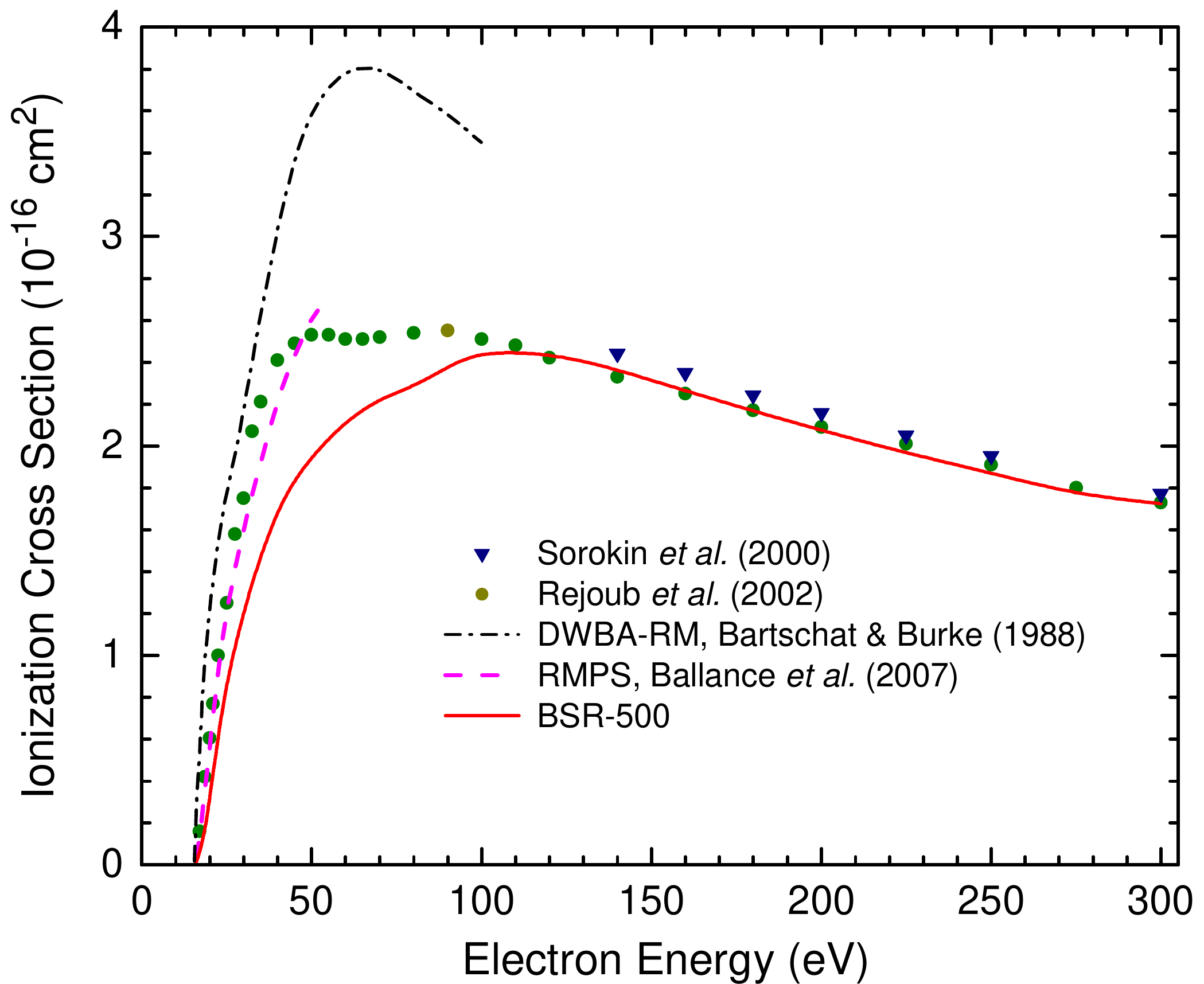}
\caption{Cross sections for electron-impact ionization of argon atoms in their ground state.
         The \hbox{BSR-500} results~\cite{BSR500} are compared with experimental data~\cite{PhysRevA.65.042713,PhysRevA.61.022723}
         and predictions from two previous calculations~\cite{0953-4075-21-17-010,0953-4075-40-3-F01}.
         }
\label{fig:Ar_ion}
\end{figure}

\begin{figure}[t]
\includegraphics[width=0.50\textwidth]{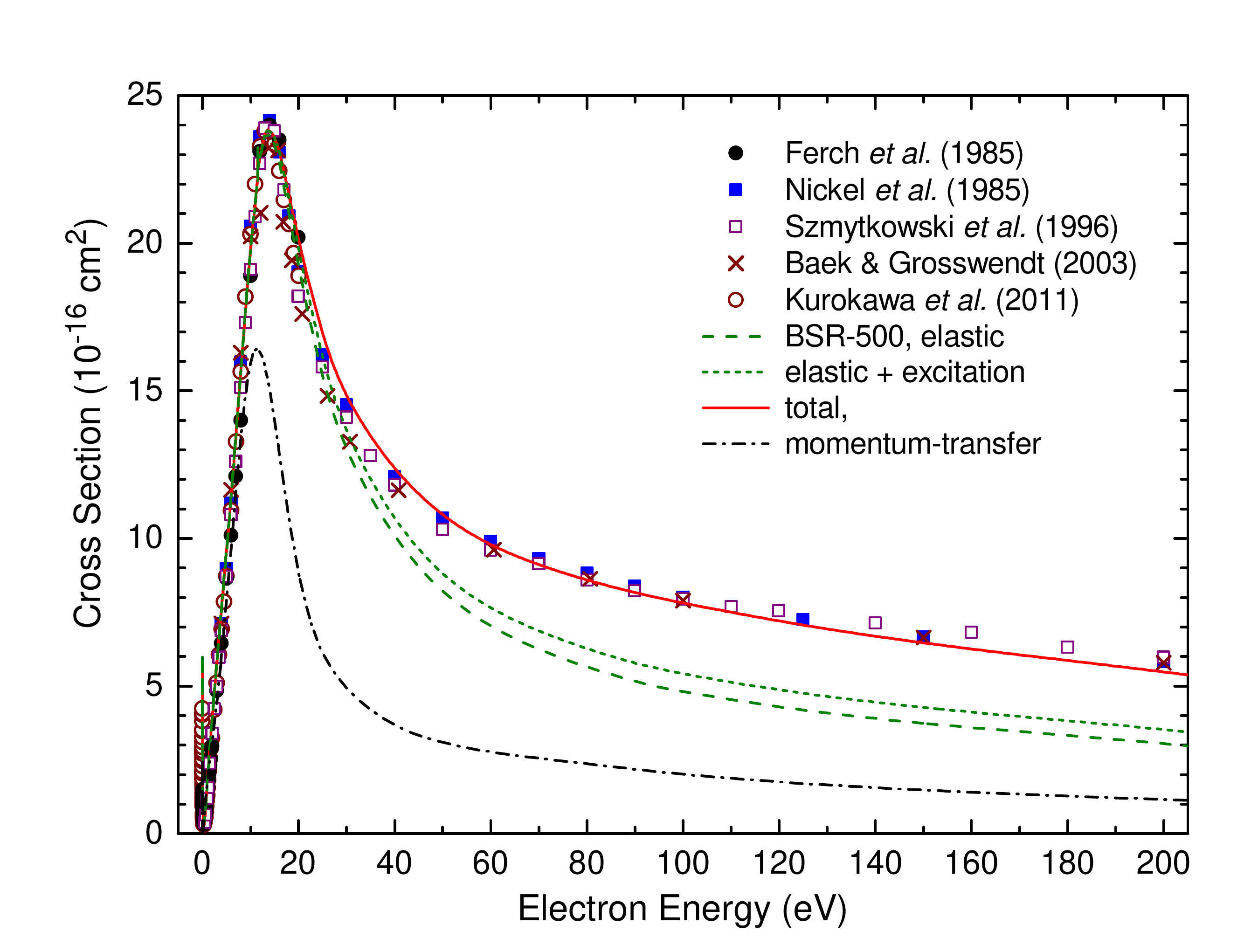}
\caption{Total cross sections for electron scattering from argon atoms in their ground state.
The \hbox{BSR-500} results~\cite{BSR500} are compared with a variety of experimental 
data~\cite{Nickel1985,Ferch1985,Szmytkowski1996,Baek2003,Kurokawa2011}.  
Also shown are the contributions
from elastic scattering alone, the sum of the elastic scattering and all excitation processes, and the momentum-transfer
cross section.}
\label{fig:Ar_total}
\end{figure}

Figure~\ref{fig:Ar_convergence} is an example of a systematic study regarding the
convergence of the close-coupling expansion~\cite{PhysRevA.89.022706}.  We see
the enormous effect of accounting for coupling to both high-lying discrete states and the ionization 
continuum on the results for these transitions from the ground state to the first four excited states.
The effect is particularly strong for the metastable $4s[3/2]_2$ and $4s'$[1/2]$_0$ states.
Somewhat surprisingly, however, the coupling effect also prevails for the excited states with
electronic angular momentum $J=1$ for incident energies at least up to 100~eV.  This fact 
suggests that simple models, such as a distorted-wave approach, would not be appropriate
until such comparatively high energies.  Semi-empirical fixes to such models, as suggested 
by Kim~\cite{PhysRevA.64.032713} with so-called ``BEf-scaling'', may help.  However, such methods are
limited to particular situations, and success is 
by no means guaranteed due to the lack of a firm theoretical foundation.

Figure~\ref{fig:Ar_4s-compare} exhibits the results again, this time as a comparison between the predictions
from the BSR-31 and \hbox{BSR-500} models with experimental data from several groups.
Without going into detail, we note that the energy dependence seen in many of the individual 
experimental data\-sets is very scattered, much more than one would expect in reality.  Consequently, the
experimental data are apparently subject to significant uncertainties, most likely due to a combination of 
statistical and systematic effects.  Based on the careful analysis of trends in the theoretical
predictions~\cite{jt642}, one would advise modelers to use the very comprehensive \hbox{BSR-500} dataset 
(state-to-state transitions between the lowest 31 states plus ionization cross sections are available
on the LXCat database~\cite{LXCat-kb})
rather than any of the few experimental data currently available.

\subsection{Ionization and Grand Total Cross Section}

Figure~\ref{fig:Ar_ion} exhibits results for ionization of the argon in its ground state.  Not surprisingly,
a hybrid model~\cite{0953-4075-21-17-010}, in which the projectile is described by a distorted wave, 
vastly overestimates the cross section for low and intermediate energies. An earlier
RMPS calculation~\cite{0953-4075-40-3-F01} does well up to incident energies of 50~eV, which is the highest energy
available from this work.  However, the trend suggests a significant increase of the predicted 
cross section for higher energies, in likely disagreement with the experimental data.  The \hbox{BSR-500} model misses
(for reasons currently unknown) the rapid increase of the cross section observed in experiment~\cite{PhysRevA.65.042713},
but agrees very well with both experimental datasets~\cite{PhysRevA.65.042713,PhysRevA.61.022723} for energies of 100~eV and above.

We finish this illustration with Fig.~\ref{fig:Ar_total}, which shows the total cross section 
(elastic + excitation + ionization).  The results obtained by the \hbox{BSR-500} model are compared with a variety
of experimental data.  Also shown is the way the various contributions make up the total, as well 
as results for the momentum-transfer cross section.  The overall agreement between experiment and BSR-500
is excellent over the entire energy range from the elastic threshold to 200~eV. 

\section{Illustration: Electron collisions with methane molecules}

Methane, CH$_4$, is a high-symmetry (tetrahedral) 10-electron molecule.
It is thus isoelectronic with neon. Song {\it et al.}~\cite{jt594}
recently completed a comprehensive review and compilation of electron
collision cross sections with methane.  It is certainly telling that, with the
exception of electron impact rotational excitations, all their recommended
cross sections are experimental. However, for some processes, 
notably electron impact electronic excitation and electron impact dissociation,
they were unable to make a firm recommendation due to the poor
quality of the available data.

Electron scattering from methane shows a number of important features.
At low energy there is a pronounced Ramsauer minimum where the cross
sections become close to zero. There are no low-lying resonances
in the system, but dissociative attachment, a process that occurs
exclusively via resonance formation, is observed around 10~eV incident energy~\cite{jt594}. 
The low-lying electronic excited states of methane
are all dissociative, and hence their excitation by electron impact is
generally assumed to lead to dissociation~\cite{zvm12}. This process
is believed to be important in plasma-assisted combustion.

$R$-matrix calculations for electron collisions with methane have been
performed by a number of groups~\cite{zvm12,npp94,jt433,vlj11,jt585}.
The results given below are largely drawn from the study performed by Brigg
{\it et al.}~\cite{jt585}. This study aimed to converge the elastic,
total, and momentum-transfer cross sections at low energy, which is the region
covering the Ramsauer minimum and, at the same time, treat the
electron impact electronic excitation/dissociation problem. 

\subsection{Elastic cross sections}

Reproducing the Ramsauer minimum is challenging because it involves
cancelation between the static interactions, which are relatively easy to reproduce,
and  polarization effects, which are not. Without
a good representation of polarization, therefore, the calculation will not give
satisfactory answers. It is actually reasonably straightforward to do
this within an SEP model~\cite{jt433}, but this model cannot treat
electronic excitation. Instead, 
Brigg {\it et al.}~\cite{jt585} tested a large number of
CC  models to see which one
converged their elastic scattering calculations.  Note, however, that 
they generally investigated eigenphase sums rather than cross sections to
test convergence. They also used the target polarizability given
by each of their models as a proxy for estimating the convergence
of their polarization potential. Brigg {\it et al.}'s final model
contained 999 states, which made the calculations computationally expensive.

\begin{figure}[t]
\includegraphics[width=0.70\textwidth]{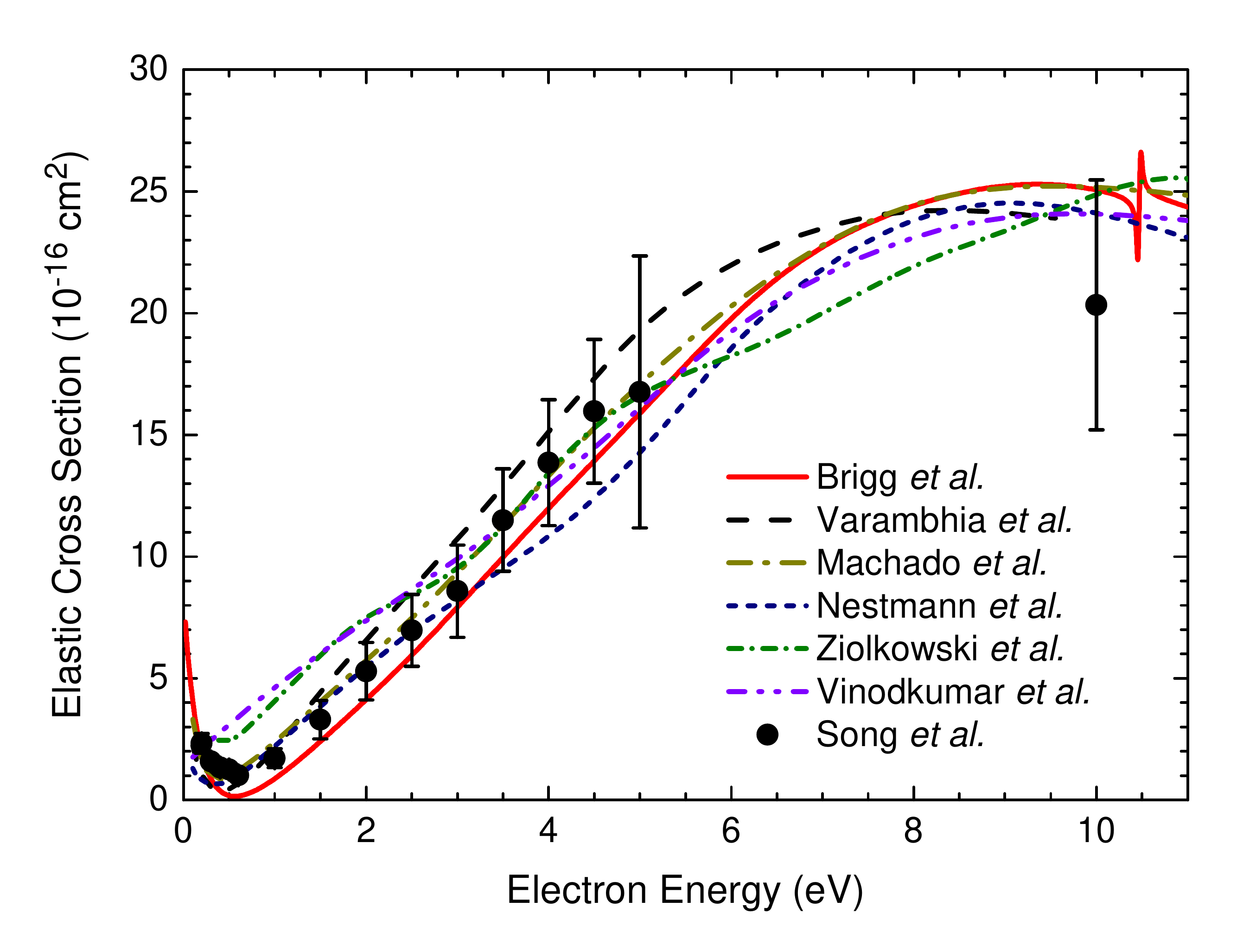}
\caption{Elastic cross sections for electron scattering from the 
methane molecule in its ground electronic state. 
The recommended (experimental) values with uncertainties given by Song {\it et al.}~\cite{jt594}.
are compared with theoretical predictions of
Brigg {\it et al.}~\cite{jt585} (RMPS),
Varambhia {\it et al.}~\cite{jt433} (SEP),
Machado {\it et al.}~\cite{Machado2002} (SEP),
Nestmann {\it et al.}~\cite{Nestmann1994} (SEP),
Ziolkowski  {\it et al.}~\cite{zvm12} \hbox{(CC-24)},
and Vinodkumar  {\it et al.}~\cite{Vinodkumar2001} \hbox{(CC-48)}.}
\label{fig:CH4_elastic}
\end{figure}

The standard molecular formulation
of the RMPS model involves single excitations from the
target wavefunction into a set of pseudo-orbitals~\cite{jt354}.  This
has proved successful for a variety of 
electron-collision problems~\cite{jt444,jt449}. However, Brigg {\it et al.} found that
for methane the model showed poor convergence and that the use of too large
a set of pseudo-states could lead to an unbalanced treatment with 
the artificial prediction of a bound CH$_4^-$ anion. Instead they 
found that it was necessary to use models closer to the 
multi-reference configuration interaction (MRCI) treatment
of the correlation problem~\cite{Werner1988}, which is frequently employed by quantum chemists. 
In this procedure
up to three electrons were excited from the target into the pseudo-states.

Figure~\ref{fig:CH4_elastic} compares a range of theoretical predictions of
the elastic cross section for electron CH$_4$ collisions. The
agreement between the RMPS/MRCI model of Brigg {\it et al.} 
and the recommended (experimental) result is good.
The SEP calculations~\cite{jt433,Machado2002,Nestmann1994} also give
reasonable agreement, while the \hbox{CC-24~\cite{zvm12}} and \hbox{CC-48~\cite{Vinodkumar2001}} calculations
do not predict a Ramsauer minimum. The latter models
do not account for coupling to the
continuum via pseudo-states and, therefore, underestimate the polarization effects.
Overall the spread of about $10\,\%$ in the values away
from the minimum is probably a reasonable estimate of the 
uncertainty in the theoretical elastic cross sections. A similar
conclusion can be drawn for the total cross section in
the low-energy regime and also for the momentum transfer cross sections;
see  Brigg {\it et al.}~\cite{jt585}.

\subsection{Excitation of nuclear motion}

Electrons collisions can excite both rotational and vibrational
motion.  Although there has been some work on electron impact vibration
excitation~\cite{Althorpe1995}, Song {\it et al.}~\cite{jt594}
concluded that these cross sections remain poorly characterized, a
conclusion that is hard to argue with.

\begin{figure}[t]
\includegraphics[width=0.70\textwidth]{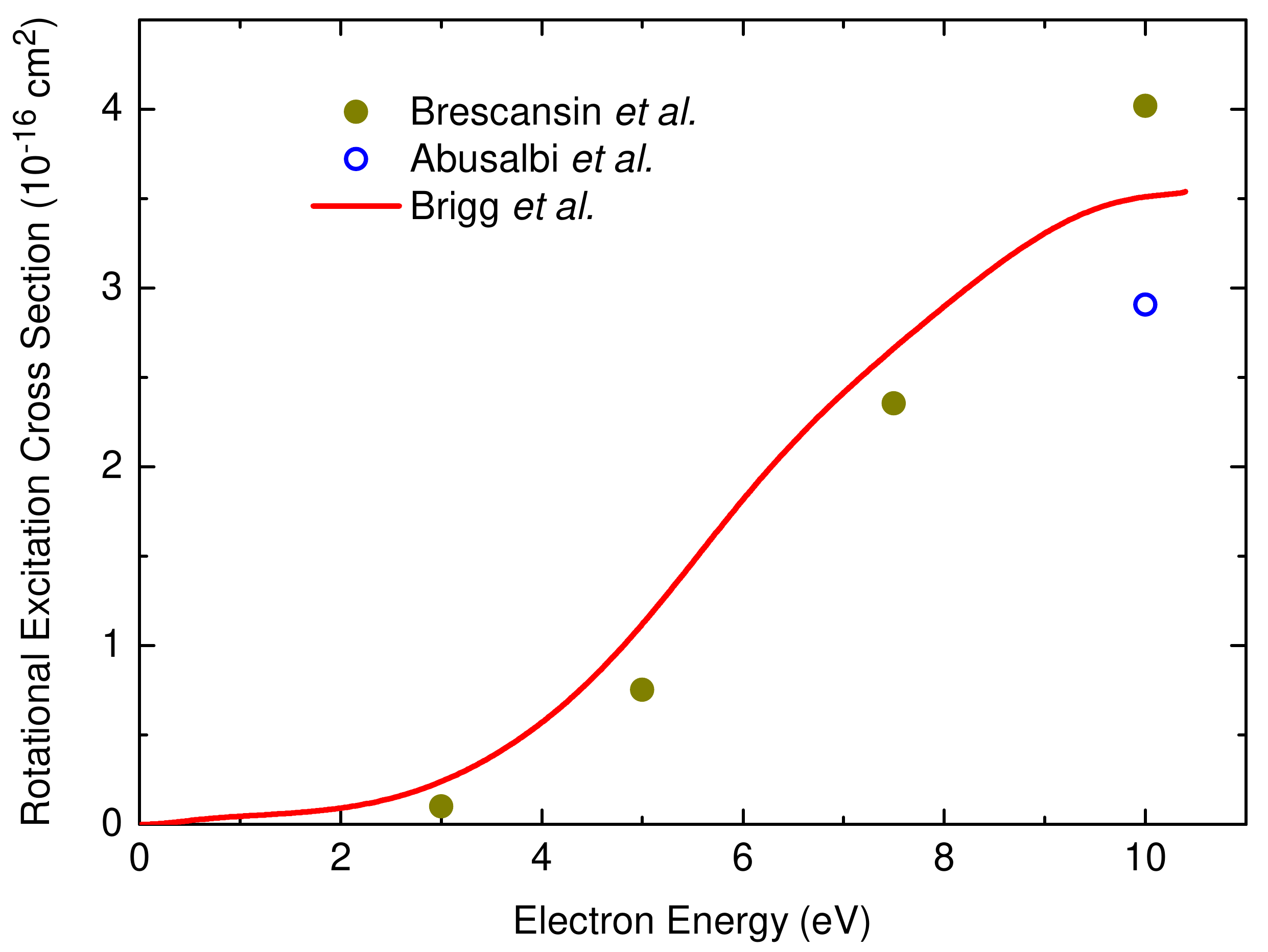}
\caption{Computed total rotational excitation cross sections for electron scattering from the 
methane molecule in its electronic ground state. The recent predictions of Brigg {\it et al.}~\cite{jt585}
are compared with earlier calculations by Brescansin {\it et al.}~\cite{Brescansin1989};
and Abusalbi {\it et al.}~\cite{Abusalbi1983}.
}
\label{fig:CH4_rot}
\end{figure}

Methane possesses neither a permanent dipole moment nor, uncommon
for a molecule, a permanent quadrupole moment. This means that
electron impact rotational excitation obeys the rather unusual
selection rule that $\Delta J \geq 3$, where $J$ is the rotational
quantum number. The resulting excitation
cross sections can be expected
to be small and dominated by short-range interactions. 
Figure~\ref{fig:CH4_rot} summarizes all currently known calculations
for the total cross section. The agreement between the individual predictions is 
reasonable, and the spread of $15\,\%$ probably approximates their uncertainty.
 
\subsection{Higher energy processes}

At higher collision energies, electron excitation, 
electron-impact dissociation (which generally
happens via excitation of dissociative electronic states~\cite{jt229}),
and ionization become important. 
Methane's low-lying electronic states are largely dissociative,
which makes cross sections for electronic excitation unimportant for plasma studies. 
Instead these excitations all contribute to the electron impact dissociation cross section, 
which remains highly uncertain for methane~\cite{jt594}.
The situation
is somewhat different for electron impact ionization.

Besides the RMPS approach~\cite{jt434}, there are a number of methods for treating
electron impact ionization of molecules.  These include ECS and TDCC,
and indeed first-principles calculations have been performed for 
electron impact ionization of methane at high energies~\cite{lmr14}.
However, experience shows that such calculations are computationally
demanding, particularly in the important intermediate-energy regime up to a 
few times the ionization threshold. Conversely, the semi-empirical
BEB method is generally found to yield excellent estimates of the
ionization cross section as a function of energy. For plasma
applications, use of the BEB procedure is likely to remain the most
practical way for the foreseeable future. Furthermore, recently
Hamilton {\it et al.}~\cite{jtNF3} have developed a practical procedure
for predicting break-up products following ionization.

\section{Conclusions and Outlook}

We have presented a brief overview of the time-independent
close-coupling method, as it is applied to calculations of cross sections
for electron collisions with atoms and molecules.  The method is based
on the (in principle) complete expansion of the total wave\-function
of the $(N\!+\!1)$-electron collision system in terms of products
describing the projectile and the $N$-electron target.  In practice, a
variety of approximations have to made, including the description of
the target structure, the level at which correlation and relativistic
effects are accounted for, and the way ionization and other
re\-arrangement processes are treated.  Consequently, estimating the
reliability of theoretical predictions has become a key aspect of any
work in this field~\cite{jt642}.  We hope that the present manuscript, together
with the references provided, will be a useful resource for the plasma
community to find relevant information when choosing which theoretical
data to incorporate in their model applications.  

At the present time
 much more robust results are obtainable
for electron collisions with atomic targets than for molecular ones.
Nevertheless, theory is still capable of providing important results for
electron collisions with molecular species~\cite{jtNF3}. For cases
involving open-shell molecules, in particular, theory remains the only source of such data.
The developments being employed to produce highly reliable electron-atom
collisions cross sections, such as the use of pseudo-states to yield converged
calculations and \hbox{$B$-spline} basis functions to allow for an accurate representation
of the continuum channels, are presently being adopted by groups working on 
electron-molecule collisions.  We expect that these efforts will 
lead to further improvement in the results of such calculations.

\section*{Acknowledgments}
This work of K.B.\ and O.Z.\ was supported by the United States National Science Foundation
under grants No.~PHY-1403245, No.~PHY-1520970, and the XSEDE super\-computer allocation 
No.~PHY-090031.


%

\end{document}